\RequirePackage{fix-cm}
\documentclass[smallcondensed,natbib]{svjour3}     
\smartqed  
\usepackage{graphicx}

\usepackage[utf8]{inputenc}
\usepackage{lineno,hyperref}
\usepackage{makecell}
\usepackage{multirow}
\usepackage{subcaption}
\usepackage[utf8]{inputenc}
\usepackage[mathscr]{euscript}
\usepackage{amssymb}
\usepackage{amsmath}
\usepackage{color}

\newtheorem{clm}{Claim}

\usepackage{tikz}
\usetikzlibrary{matrix}
\usetikzlibrary{positioning}
\usetikzlibrary{graphs}

\definecolor{myblue}{RGB}{80,80,160}
\definecolor{mygreen}{RGB}{80,160,80}
\usetikzlibrary{positioning,chains,fit,shapes,calc,arrows}

\tikzset{ 
	mylabel/.style={draw=none,fill=none},
    table/.style={
        matrix of nodes,
        row sep=-\pgflinewidth,
        column sep=-\pgflinewidth,
        nodes={
            rectangle,
            draw=black,
            align=center
        },
        minimum height=2em,
        text depth=0.5ex,
        text height=2ex,
        nodes in empty cells,
        row 1/.style={
            nodes={
                fill=mygreen,
                text=white,
                font=\bfseries
            }
        },
        column 1/.style={
            nodes={text width=3em,
            fill=myblue,
                text=white,
                font=\bfseries}
        }
    }
}

\tikzset{ 
    table2/.style={
        matrix of nodes,
        row sep=-\pgflinewidth,
        column sep=-\pgflinewidth,
        nodes={
            rectangle,
            draw=black,
            align=center
        },
        minimum height=1.5em,
        text depth=0.5ex,
        text height=2ex,
        nodes in empty cells,
        row 1/.style={
            nodes={
                fill=black,
                text=white,
                font=\bfseries
            }
        },
        column 1/.style={
             nodes={
                fill=black,
                text=white,
                font=\bfseries
            }
        }        
    }
}

\journalname{Machine Learning}
\begin{document}

\title{Algorithm Selection for Collaborative Filtering: the influence of graph metafeatures and multicriteria metatargets\thanks{This work is financed by the Fundação para a Ciência e a Tecnologia through grant number SFRH/BD/117531/2016 also supported by CNPq and FAPESP, Brazilian funding agencies.}
}

\titlerunning{Algorithm Selection for Collaborative Filtering}        

\author{Tiago Cunha         \and
        Carlos Soares 		\and \\
        André C. P. L. F. de Carvalho 
}


\institute{Tiago Cunha, Carlos Soares \at
              Faculdade de Engenharia da Universidade do Porto, Portugal \\
              \email{\{tiagodscunha,csoares\}@fe.up.pt}           
           \and
           André C. P. L. F. de Carvalho \at
           Instituto de Ciências Matemáticas e de Computação - Universidade de São Paulo, Brazil \\
           \email{andre@icmc.usp.br}           
}

\date{Received: date / Accepted: date}

\maketitle

\begin{abstract}
To select the best algorithm for a new problem is an expensive and difficult task. However, there are automatic solutions to address this problem: using Metalearning, which takes advantage of problem characteristics (i.e. metafeatures), one is able to predict the relative performance of algorithms. In the Collaborative Filtering scope, recent works have proposed diverse metafeatures describing several dimensions of this problem. Despite interesting and effective findings, it is still unknown whether these are the most effective metafeatures. Hence, this work proposes a new set of graph metafeatures, which approach the Collaborative Filtering problem from a Graph Theory perspective. Furthermore, in order to understand whether metafeatures from multiple dimensions are a better fit, we investigate the effects of comprehensive metafeatures. These metafeatures are a selection of the best metafeatures from all existing Collaborative Filtering metafeatures. The impact of the most representative metafeatures is investigated in a controlled experimental setup. Another contribution we present is the use of a Pareto-Efficient ranking procedure to create multicriteria metatargets. These new rankings of algorithms, which take into account multiple evaluation measures, allow to explore the algorithm selection problem in a fairer and more detailed way. According to the experimental results, the graph metafeatures are a good alternative to related work metafeatures. However, the results have shown that the feature selection procedure used to create the comprehensive metafeatures is is not effective, since there is no gain in predictive performance. Finally, an extensive metaknowledge analysis was conducted to identify the most influential metafeatures.


\end{abstract}

\section{Introduction}

The algorithm selection problem has been frequently addressed with Metalearning approaches (MtL)~\citep{Vilalta1999,Hilario2000,Brazdil2003,Prudencio2004,Smith-Miles2008a,Gomes2012,Lemke2013,Rossi2014}. This technique finds the mapping between problem-specific characteristics (i.e. metafeatures) and the relative performance of learning algorithms (i.e. metatarget)~\citep{Brazdil:2008:MAD:1507541}. This mapping, provided as a Machine Learning model (i.e. metamodel), can then be used to predict the best algorithms for a new problem. This task is organised into a baselevel and a metalevel. The baselevel refers to the learning task for which recommendations of algorithms are made, which, in this case, is Collaborative Filtering (CF). The metalevel refers to the learning task which studies the mapping between metafeatures and algorithm performance. In this work, the metalevel is addressed as a Label Ranking task. 

Several algorithm selection approaches have been recently proposed for CF~\citep{Adomavicius2012,Ekstrand2012,Griffith2012,Matuszyk2014,Cunha2016,Cunha2017,Cunha2018128}. In spite of their contribution to important advances in the area, there are still limitations that need to be addressed. These limitations are mainly related to the metafeatures and the metatarget, which are the focus of this work. 

The main limitation regarding metafeatures is that most approaches only describe the recommendation problem using descriptive characteristics of the rating matrix and estimates of performance on samples (i.e. landmarkers), overlooking a wide spectrum of other possibilities. Furthermore, existing papers typically perform a limited comparison between the proposed metafeatures and the ones proposed in other studies. Additionally, there is a lack of studies combining metafeatures from multiple domains in a single collection and validating their individual and combined merits in the same experimental setup. 

Regarding the metatarget, the limitation lies mainly in the fact that the best algorithms per dataset are considered using only one evaluation measure at a time. Hence, to do algorithm selection according to additional measures, it is necessary to replicate the experimental procedure for each measure, represented as a different metatarget. Beyond the eficiency issues, this process is not ideal since it leads to limited and measure-dependent metaknowledge. Hence, an alternative must be found, ideally in a way which allows for a multitude of evaluation measures to be used simultaneously.

This work proposes solutions for these limitations. These are evaluated in a comprehensive experimental study. and provide the following novel contributions to the problem of CF algorithm selection: 

\begin{itemize}
\item \textbf{Graph metafeatures: } By modelling the CF problem as a bipartite graph, one is able to use an alternative way to describe the relationships between users and items. For such, this work proposes metafeatures based on Graph Theory~\citep{west2001introduction,godsil2013algebraic} and adopts aspects from systematic and hierarchical metafeature extraction processes~\citep{Cunha2016,Cunha:2017:MCF:3109859.3109899}.
\item \textbf{Comprehensive metafeatures: } A set of metafeatures obtained by taking advantage of metafeatures from multiple domains: Rating Matrix~\citep{Cunha2016}, Landmarkers~\citep{Cunha2017} and Graph metafeatures. 
\item \textbf{Multicriteria metatarget: } The metatarget is obtained by aggregating the rankings of algorithms produced by multiple evaluation measures. We adapt Pareto-Efficient rankings~\citep{Ribeiro2013} to CF algorithm selection.
\end{itemize}

This document is organised as follows: Section~\ref{sec:rw} introduces the related work on CF, MtL and algorithm selection for CF, while Sections~\ref{sec:main} and ~\ref{sec:metatargets} presents the main contributions, respectively. In Section~\ref{sec:setup}, the empirical setup is presented and Sections~\ref{sec:preliminary} and ~\ref{sec:results} discuss both the preliminary analysis and the empirical results, respectively. Section~\ref{sec:conclusions} presents the conclusions and directions for future work.

\section{Related Work}\label{sec:rw}

\subsection{Collaborative Filtering}\label{sub:cf}

CF recommendations are based on the premise that a user will probably like the items favoured by a similar user. Thus, CF employs the feedback from each individual user to recommend items to similar users~\citep{Yang2014}. The feedback is a numeric value, proportional to the user's interest in an item. Most feedback is based on a rating scale, although  variants, such as like/dislike binary responses, are also employed. The data structure is a rating matrix $R$. It is usually described as $R ^{|U| \times |I|}$, representing a set of users $U$ and a set of items $I$. Each element of this matrix is the feedback provided by each user for each item. 

CF algorithms can be organised into memory- and model-based~\citep{Bobadilla2013a}. Memory-based algorithms apply heuristics to a rating matrix to extract recommendations, whereas model-based algorithms induce a predictive model from this matrix, which can later be used for future recommendations. Most memory-based algorithms adopt Nearest Neighbour strategies, while model-based are mostly based on Matrix Factorization methods~\citep{Yang2014}. Further discussion regarding CF algorithms is available elsewhere~\citep{Yang2014}.

The evaluation of Recommender Systems (RSs) is usually performed by procedures that split the dataset into training and test subsets (using sampling strategies, such as k-fold cross-validation~\citep{Herlocker2004}) and assess the performance of the induced model on the test dataset. Different evaluation metrics can be used~\citep{Lu2012}. The evaluation measures used depend on the type of prediction: for ratings of the items, error measures like Mean Absolute Error (MAE) or Root Mean Squared Error (RMSE) are used; for binary relevance, Precision/Recall or Area Under the Curve (AUC) are used; finally, a common measure for rankings of items is the Normalised Discounted Cumulative Gain (NDCG).

\subsection{Algorithm selection using Metalearning}\label{sub:mtl}

The algorithm selection problem has been conceptualised in 1976 by Rice~\citep{DBLP:journals/ac/Rice76}. It involves the following search spaces: the problem space $P$, the feature space $F$, the algorithm space $A$ and the performance space $Y$. These refer respectively to problem instances, features, algorithms and evaluation measures.  The problem is formulated as: for a given instance  $ x \in P $, with features $ f(x) \in F $, find the selection mapping $ S(f(x)) $ into space $ A $, such that the selected algorithm  $ \alpha \in A $ maximises the performance mapping $ y(\alpha(x)) \in Y $\citep{DBLP:journals/ac/Rice76}. 

One of the main challenges in MtL is to define which metafeatures effectively describe how a problem matches the bias of an algorithm~\citep{Brazdil:2008:MAD:1507541}. The MtL literature often divides the metafeatures into three main groups~\citep{Serban2013,Vanschoren2010}: 

\begin{itemize}
\item \textbf{Statistical and/or information-theoretical} describe the dataset using a set of measures from statistics and information theory. Examples include simple measures, like the number of examples and features, as well as more advanced measures, like entropy, skewness and kurtosis of features and even mutual information and correlation between features; 
\item \textbf{Model-based} metafeatures are made of properties extracted from models induced from a dataset. As an example, if a decision tree induction algorithm is applied to a dataset, one model-based metafeature could be
the number of leaf nodes in the decision tree. The rationale is that there is a relationship between model characteristics and algorithm performance that cannot be directly captured from the dataset. 
\item \textbf{Landmarkers} are fast estimates of the performance of an algorithm on a given dataset. Since these estimates are used as metafeatures, it is important that they are computationally much faster than applying the algorithm to the dataset (e.g. using hold-out to estimate performance). Two different types of landmarkers can be obtained by 1)  applying fast and/or simplified algorithms on complete datasets (e.g. a decision stump can be regarded as a simplified version of a decision tree) ; 2) applying conventional algorithms to a sample extracted from a dataset, also known as \textbf{subsampling landmarkers}~\citep{Brazdil:2008:MAD:1507541} (e.g. applying a decision tree induction algorithm to a sample extracted from a dataset). 
\end{itemize}

Recently, a systematic metafeature framework~\citep{Pinto2016} has been proposed to simplify the process of designing metafeatures for a MtL task. The framework requires three main elements: objects $o$ (e.g. numeric variables), functions $f$ (e.g. correlation) and post-functions $pf$ (e.g. average). In order to generate a single metafeature value, the metafeature extraction procedure applies each function to all possible set of compatible objects (e.g., correlation between every pair of numeric variables). This yields multiple values,  and a post-function is applied to those values to obtain a metafeature. The metafeatures created using this framework are represented as: $\{o\}.\{f\}.\{pf\}$. One important property of this framework is recursiveness. As a result, the outcome of an inner level (IL) application of the framework can be used as the result of an outer level (OL) function. Formally:

$
\{\textrm{OL-}o\}.\{\textrm{OL-}f\}.\{\textrm{OL-}pf\} = 
\{\textrm{OL-}o\}.
\Big[\{\textrm{IL-}o\}.\{\textrm{IL-}f\}.\{\textrm{IL-}pf\}\Big].
\{\textrm{OL-}pf\}
$

\subsection{Metalearning for Collaborative Filtering}\label{sub:mtl_cf}

Recently, a few MtL approaches were proposed to the problem of selecting CF algorithms. Two types of metafeatures have been used for that purpose, statistical and/or information-theoretical and subsampling landmarkers. In the rest of this document, we will use the terms metafeature and meta-approach to refer to the descriptors and MtL approaches, respectively.

\paragraph{Statistical and/or information-theoretical} Existing studies have made arbitrary choices in the development of metafeatures~\citep{Adomavicius2012,Ekstrand2012,Griffith2012,Matuszyk2014}, A systematic approach to the design of metafeatures for CF was proposed recently~\citep{Cunha2016}. These metafeatures  describe the rating matrix using the systematic framework summarised earlier. It analyses an extensive combination of a set of objects $o$ (rating matrix $R$, and its rows $U$ and columns $I$), a set of function $f$ (original ratings (\textit{ratings}), number of ratings (\textit{count}), mean rating value (\textit{mean}) and sum of ratings (\textit{sum})), and a set of post-functions $pf$ (maximum, minimum, mean, standard deviation, median, mode, entropy, Gini index, skewness and kurtosis). This class of metafeatures will be identified as RM from this point onward. 

\paragraph{Subsampling Landmarkers} A single approach uses this type of metafeatures in the CF scope~\citep{Cunha2017}. These data characteristics are obtained by assessing the performance of the CF algorithms on random samples of the datasets. These estimates are combined to create different metafeatures. Performance is estimated using different evaluation measures, which leads to a set of metafeatures for each measure. Although the work studied different landmarking perspectives (i.e. relative landmarkers~\citep{Furnkranz2002}), which manipulate the values in different ways in order to properly explore the problem, no significant gain of performance was obtained. Therefore, this work considers simply the performance values as metafeatures. The format used to describe these metafeatures is: \textit{algorithm.evaluation measure}. This class of metafeatures is referred to as SL in the remainder of this document.

\section{Graph metafeatures}\label{sec:main}

Given that CF's rating matrix can be regarded as a (weighed) adjacency matrix, it means that a CF problem can be represented as a graph. It is our belief that the extraction of new metafeatures using this graph representation can provide new information not captured by other meta-approaches. Among other benefits, it allows not only to model but also to describe the problem in more detail. Thus, the main motivations for this new approach are two-fold:

\begin{itemize}
\item \textbf{Data structure compatibility:} The rating matrix data can be correctly described using a bipartite graph. For such, it can be assumed that rows and columns refer to independent sets of nodes and that the feedback values stored within the matrix are represented as edge between nodes.
\item \textbf{Neighbourhood characterisation:} Metafeatures that characterize users in terms of their neighbourhood have been used before in algorithm selection for CF~\citep{Griffith2012}. The approach is capable of creating user-specific metafeatures, responsible to describe a user by statistics of its neighbours. However, it is not able to generate metafeatures which represent all neighbourhoods in a dataset. Hence, if the problem is represented by a graph, extracting complex neighbourhood statistics becomes easy.
\end{itemize}

\begin{figure}[!ht]
    \centering
    \begin{subfigure}[b]{0.4\textwidth}
        \begin{tikzpicture}
            \matrix (first) [table,text width=2em]
            {
            |[mylabel]| & ${\boldsymbol{i}}_{\boldsymbol{1}}$   & ${\boldsymbol{i}}_{\boldsymbol{2}}$ & ${\boldsymbol{i}}_{\boldsymbol{3}}$ \\
            $\boldsymbol{u}_{\boldsymbol{1}}$   	& ${\boldsymbol{5}}$ & ${\boldsymbol{3}}$ & ${\boldsymbol{4}}$ \\
            $\boldsymbol{u}_{\boldsymbol{2}}$    	& ${\boldsymbol{4}}$ & $\dots$ & ${\boldsymbol{2}}$  \\
            $\boldsymbol{u}_{\boldsymbol{3}}$   	& $\dots$ & ${\boldsymbol{3}}$  & ${\boldsymbol{5}}$  \\
            };
        \end{tikzpicture}
        \caption{Rating Matrix.}
        \label{fig:rm}
    \end{subfigure}
    ~ 
    \begin{subfigure}[b]{0.4\textwidth}
    	\begin{tikzpicture}[thick,
                fsnode/.style={},
                ssnode/.style={},
                every fit/.style={ellipse,draw,inner sep=1pt,text width=1cm},
                ->,shorten >= 1pt,shorten <= 1pt
                ]

        \begin{scope}[start chain=going below,node distance=3mm]
        \foreach \i/\xcoord/\ycoord in {1/6/8,2/5/1,3/-4/7}
        \node[fsnode,on chain,label=left:$u_{\i}$] (f\i) {};
        \end{scope}

        \begin{scope}[xshift=3cm,start chain=going below,node distance=3mm]
        \foreach \i/\xcoord/\ycoord in {1/0/3,2/1/4,3/-2/1}
        \node[ssnode,on chain,label=right:$i_{\i}$] (s\i) {};
        \end{scope}

        \node [myblue,fit=(f1) (f3),label=above:$U$, line width=0.5mm] {};
        \node [mygreen,fit=(s1) (s3),label=above:$I$, line width=0.5mm] {};

        \draw (f1) edge[bend left=45] node[label={[xshift=0cm,yshift=-0.05cm]5}]{} (s1) ;
        \draw (f1) edge[bend left=30] node[label={[xshift=-0.1cm,yshift=-0.05cm]3}]{} (s2);
        \draw (f1) edge[bend left=15] node[label={[xshift=-0.1cm,yshift=-0.5cm]4}]{}  (s3);
        \draw (f2) edge[bend left=15] node[label={[xshift=0.02cm,yshift=-0.45cm]4}]{} (s1);
        \draw (f2) edge[bend right=15] node[label={[xshift=0.04cm,yshift=-0.2cm]2}]{} (s3);
        \draw (f3) edge[bend right=30] node[label={[xshift=-0.05cm,yshift=-0.5cm]3}]{} (s2);
        \draw (f3) edge[bend right=45] node[label={[xshift=0.03cm,yshift=-0.6cm]5}]{} (s3);
        \end{tikzpicture}
        \caption{Bipartite Graph.}
        \label{fig:gr}
    \end{subfigure}
    \caption{Toy example for two different CF representations.}\label{fig:toy_example}
\end{figure}
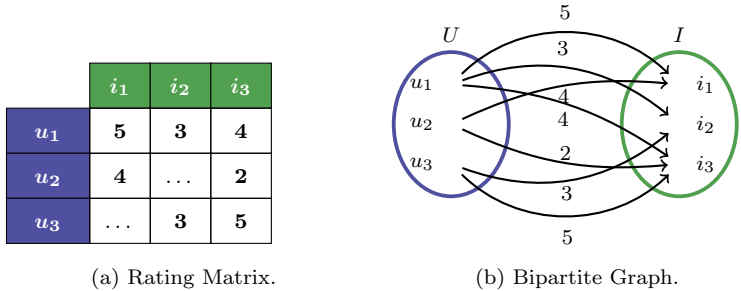

As a result, this study models the problem as a bipartite graph $G$, whose nodes $U$ and $I$ represent users and items, respectively. The set of edges $E$ connects elements of the two groups and represent the feedback of users regarding items. The edges can be weighted, hence representing preference values (i.e. ratings). Figure~\ref{fig:toy_example} presents an example with the two representations for the same CF problem.

The proposed graph meta-approach is based on Graph Theory~\citep{west2001introduction,godsil2013algebraic}. Although the literature provides several functions for graph characterisation that can be used for this purpose, they have a major limitation: the characteristics describe the graph at a high-level, which limits the information to be extracted. For instance, the amount of information available in measures such as the number of nodes or edges is limited for our purpose. 

To deal with this limitation, we use the systematic metafeature extraction~\citep{Pinto2016} and hierarchical decomposition of complex data structures~\citep{Cunha:2017:MCF:3109859.3109899} approaches for metafeature design. It is important to notice that since we were unable to find any graph-based metafeatures in the literature, we have adopted an exploratory approach: this means that we use as many graph characterisation features as possible and then try to identify which ones are informative. Then, we propose metafeatures extracted from graphs at different levels: 

\begin{itemize}
\item \textbf{Graph-level} properties that describe the graph in a high level perspective; 
\item \textbf{Node-level} characteristics relating nodes through their edges relationships; 
\item \textbf{Pairwise-level} properties obtained by node pairwise comparisons;
\item \textbf{Subgraph-level} characteristics summarising relevant subgraphs. 
\end{itemize}

\subsection{Graph-level}

When trying to propose metafeatures for a complex structure, it is common to consider high level characteristics first. Although in the context of algorithm selection this is not typically effective~\cite{Cunha:2017:MCF:3109859.3109899}, it is nevertheless important to verify it. Hence, at this level, only one object is considered for metafeature extraction: the whole bipartite graph $G$, which
can be directly characterised through several Graph Theory measures~\citep{west2001introduction,godsil2013algebraic}. This work selects a subset of potentially useful characteristics to be used as metafeatures. These are: 

\begin{equation}
G.\{edge\ density, girth, order, size, radius\}.\varnothing
\end{equation}

The functions refer, respectively, to the ratio of the number of existing edges over the number of possible edges, length of the shortest circle, number of nodes, number of edges and the smallest maximum distance between the farthest nodes of the graph. The formalisation of these functions lie outside the scope of this work.\footnote{The interested reader may find more information in the graph theory literature~\citep{west2001introduction,godsil2013algebraic}.}
Since these functions return a single value, no metafeature used at this level requires post-processing. This is represented by the symbol $\varnothing$.

\subsection{Node-level}

In this level we argue that since nodes represent the main entities in the graph, it is potentially beneficial to extract characreristics which represent them and their edges on a global perspective. Specifically in this case, where two clearly well defined sets of nodes exist (i.e. users and items), it is important to find suitable characteristics for each one. If one is able to properly characterize the users through their relationships to items (and vice-versa), then hopefully we will be able to find metafeatures able to represent the way CF operates: new items are recommended based on the preferences of users with similar tastes. 

Hence, Node-level metafeatures use three different objects: the graph $G$, the set of users $U$ and the set of items $I$. These consider the entire graph and each subset independently. This separation of concepts allows a more extensive analysis and to understand whether the different subsets of nodes hold different degrees of importance for the MtL problem. For instance, if we find that metafeatures related to the users are not informative, then this presents interesting insights to the algorithm selection problem. However, if we considered all nodes, we are unable to make such analysis. The functions used at this level describe the nodes through their edge relationships. We select a wide variety of functions which are suitable to describe bipartite graphs:

\begin{itemize}
\item \textbf{Alpha centrality: } Bonacich's alpha centrality~\citep{BONACICH2001191};
\item \textbf{Authority score: } Kleinberg's authority score~\citep{Kleinberg:1999:ASH:324133.324140};
\item \textbf{Closeness centrality:} the inverse of the average length of the shortest paths to/from all the other nodes in the graph;
\item \textbf{Constraint: } Burt's constraint score~\citep{doi:10.1086/421787};
\item \textbf{Coreness: } the coreness of a node is $k$ if it belongs to the $k$-core (maximal subgraph in which each node has at least degree $k$) but not to the ($k+1$)-core.
\item \textbf{Degree: } the number of adjacent edges;
\item \textbf{Diversity: } the Shannon entropy of the weights of a node's incident edges;
\item \textbf{Eccentricity: } shortest path from the farthest node in the graph;
\item \textbf{Eigenvector Centrality score: } the values of the first eigenvector of the adjacency matrix;
\item \textbf{Hub score:} Kleinberg’s hub centrality score~\citep{Kleinberg:1999:ASH:324133.324140};
\item \textbf{KNN:} average nearest neighbour degree;
\item \textbf{Neighbours;} amount of adjacent nodes in a graph;
\item \textbf{Local Scan:} average edge weights;
\item \textbf{PageRank:} Google's PageRank score per node.
\item \textbf{Strength:} sum of adjacent edges weights.
\end{itemize}

Since the application of these functions to the nodes of a graph return a set of values, these values must be aggregated to have a single value for the metafeature. To do so, this work employs post-processing functions $pf$, which return the following single values: mean, variance, skewness and entropy. These functions, based on statistical univariate analysis (central tendency, dispersion and shape) and Information Theory, have performed well in other recommendation metafeatures~\citep{Cunha:2017:MCF:3109859.3109899}. These metafeatures can be formally described as:

\begin{equation}
\begin{aligned}
\{G,U,I\}.\{alpha, authority, closeness, constraint, coreness, degree, diversity, \\eccentricity, eigenvector, hub, knn, neighbours, scan, PageRank, strength\}.\{pf\}
\end{aligned}
\end{equation}

\subsection{Pairwise-level}

Having exhausted the ability to characterize nodes by their explicit edge relationships, one must find alternative ways to explore implicit patterns. A methodology which proved to be successful in other algorithm selection domains~\cite{Cunha:2017:MCF:3109859.3109899} performs pairwise comparisons of simpler elements in the complex data structure and aggregates its values to create a global score to characterize the entire structure. These comparisons allow to understand whether there are important relationships among said elements which represent overall patterns. 

Hence, the pairwise metafeatures designed in this level are based on the comparison among all pairs of nodes. Due to the complexity of the data structure, the pairwise-level defines 2 layers - inner (IL) and outer (OL) - which we present next.

\subsubsection{Inner Layer (IL)}

The IL, responsible for node comparison, applies pairwise comparison functions to all pairs of nodes $n_i, n_j$. The output is stored in the specific row $i$ and column $j$ of a  IL matrix, used to keep records of pairwise comparisons. Figure~\ref{fig:pairwise} presents such data structure, with rows and columns referring to the same set of nodes. 

\begin{figure}[!h]
\centering
\begin{tikzpicture}
\matrix (first) [table2,text width=2em]
{
|[mylabel]| & ${\boldsymbol{n}}_{\boldsymbol{1}}$   & ${\boldsymbol{n}}_{\boldsymbol{2}}$ & ${\boldsymbol{n}}_{\boldsymbol{3}}$ \\
$\boldsymbol{n}_{\boldsymbol{1}}$   	& $\dots$ &$\dots$   &$\dots$   \\
$\boldsymbol{n}_{\boldsymbol{2}}$    	& $\dots$ &$\dots$   &$\dots$   \\
$\boldsymbol{n}_{\boldsymbol{3}}$   	& $\dots$ &$\dots$   &$\dots$   \\
};
\end{tikzpicture}
\caption{IL matrix for all nodes $g \in G$.}
\label{fig:pairwise}
\end{figure}

The functions used to perform pairwise comparisons are:
\begin{itemize}
\item \textbf{Similarity:} the number of common neighbours divided by the number of nodes that are neighbours of at least one of the two nodes being considered;
\item \textbf{Distance:} length of the shortest paths between nodes.
\end{itemize}

The post-processing functions used in this layer are the matrix post-processing functions ($mpf$). The sum, mean, count and variance functions are applied to each matrix row (alternatively, given the symmetry in the IL matrix, could be applied to each column). The output is a set of summarised comparison values for each function. Such values are submitted to the OL to obtain the final metafeatures.

\subsubsection{Outer Layer (OL)}

The OL takes advantage of the recursiveness in the systematic metafeature framework. It does so by using the same objects as used in the Node-level: $G$,$U$,$I$. Each of these sets of nodes are separately submitted to the IL to obtain the actual node comparison scores. This means that effectively we perform 3 IL operations. Finally, the values returned by each set of nodes are aggregated to create the final metafeatures using the same post-processing functions as before: mean, variance, skewness and entropy. The formalization of the metafeatures in this level is (refer to Section~\ref{sub:mtl} for interpretation of the recursive notation used):

\begin{equation}
\{G,U,I\}.\Big[\{g_i/g_j,u_k/u_l,i_m/i_n\}.\{similarity, distance\}.\{mpf\}\Big].\{pf\}
\end{equation}

\subsection{Subgraph-level}

So far, we have described measures that characterize the whole graph or very small parts of it (nodes and pairs of nodes). However, a graph may contain parts that have very specific structures, which are different from the rest (e.g. the most popular items will define a very dense subgraph). Therefore, it is important to include metafeatures that provide information about those subgraphs. Hence, the metafeatures at this level split the graph into relevant subgraphs, describes each one with specific functions and aggregates the final outcome to produce the metafeature. Once again, due to complexity, we define one IL and one OL. We start by describing how a subgraph is characterized in the IL and move to the OL afterwards.

\subsubsection{Inner Layer (IL)}

The IL assumes the existence of a subgraph. Our proposal is to use Node-level metafeatures to describe it. We could also include the Pairwise-level metafeatures also in this scope. However, due to the high computational resources required we have discarded them at this stage. Since the outcome is a metafeature value for each node in the subgraph, the values necessary to describe the overall subgraph must be aggregated. In order to deal with this issue, the mean, variance, skewness and entropy $pf$ functions are used. 

\subsubsection{Outer Layer (OL)}

The OL is responsible to create the subgraphs to be provided to the IL. The subgraphs characterised here are:

\begin{itemize}
\item \textbf{Communities:} obtained using the Louvain's community detection~\citep{DBLP:journals/corr/abs-0803-0476} algorithm, which operates by multilevel optimisation of modularity;
\item \textbf{Components:} subgraphs of maximal strongly connected nodes of a graph.
\end{itemize}

After providing each community and component to the IL, one must once again aggregate the results. This is necessary to obtain a fixed-size description of the communities and components  that characterizes a varying number of its subgraphs. These metafeatures can be formally defined as:

\begin{equation}
\{communities, components\}.\Big[\{subgraph\}.\{Node-level\}.\{pf\}\Big].\{pf\}
\end{equation}

\section{Multicriteria Metatarget}\label{sec:metatargets}

MtL focuses mainly on which are the most informative metafeatures to predict the best algorithms~\citep{Adomavicius2012,Ekstrand2012,Griffith2012,Matuszyk2014,Cunha2016,Cunha2017,Cunha2018128}. However, the way the best algorithms are selected to build the metatarget is usually simplified: a specific evaluation metric is selected and used to assess the performance of a set of algorithms on a specific dataset. Then, the best algorithm according to that specific dataset is used as its metatarget. 

The main problem with this approach is that a single evaluation measure is usually not enough to properly and completely characterize the performance of an algorithm. In fact, this has been identified as a particularly important issue in the RS scope~\citep{Herlocker2004,Gunawardana2009a,Ciordas2010}, as multiple, sometimes conflicting, measures are equally important (e.g. precision and recall). Hence, it makes sense that any MtL approach for RS methods must analyse the algorithm selection problem, while taking into account the inputs of multiple evaluation measures to create a multicriteria metatarget. 

This section describes our proposal to tackle this issue: the multicriteria metatarget. It is important to notice that unlike earlier works which considered only the best algorithm per dataset to build the metatarget~\citep{Cunha2016,Cunha2017,Cunha2018128}, this work builds upon a recent work which has shown the importance of using rankings of algorithms~\citep{Cunha2018}. Hence, our multicriteria metatarget procedure takes into account algorithm rankings provided by various evaluation measures to create a multicriteria ranking of algorithms. 

Before dwelling in the inner workings of the procedure, let us assign proper notation. Let us assume the following concepts: consider $D$ as the group of CF datasets, $A$ as an ordered collection of CF algorithms and $M$ as the set of evaluation measures. To create the metatargets, first every dataset $d_i \in D$ is subjected to all algorithms $a_j \in A$ to create recommendation models. Afterwards, every model is evaluated using a specific evaluation measure $m_l \in M$ in order to obtain a performance $p_{m_l} (a_j|d_i)$, which characterizes how good the model is for that problem accordingly to the scope the evaluation measure assesses. Then, for every $d_i$ and measure $m_l$, the performance values $p_{m_l} (a_j|d_i)$ are sorted with decreasing degree of importance to create an array of algorithm rankings $\theta$. This ranking refers to positions in the ordered collection of algorithms $A$, meaning that $\theta \in [1,|A|]$. Notice that $\theta$ can also be regarded as a sequence of pairs of algorithms and rankings positions: $\big(<a_j,\theta_j>\big)_{j=1}^{|A|}$. Formally, the ranking metatarget is:

\begin{equation}
mt(d_i,m_l) = SORT\big(p_{m_l} (a_j|d_i)\big)_{j=1}^{|A|}
\end{equation}

The problem addressed here lies in the cases where more than one evaluation measure $m_l$ must be used to create the multicriteria metatarget. To do so, we adapt Pareto-Efficient Rankings~\citep{Ribeiro2013}, originally proposed to create a single ranked lists of items using rankings predicted by different recommendation algorithms. First, let us inspect the original rationale: consider a User-Interest space, which is used to represent the preferences each algorithm defines for multiple Items for a specific User. This space is used to define Pareto frontiers, which in turn allows to create multicriteria rankings of Items considering the inputs of multiple Algorithms. Notice the obvious parallelism to our problem: if we consider that User, Algorithm and Item concepts are now represented as Dataset, Evaluation Measure and Algorithm, then the task can be similarly expressed: 

\begin{clm}
For every \{User/Dataset\}, create a ranking of \{Items/Algorithms\} which considers the preferences of multiple \{Algorithms/Evaluation Measures\}.
\end{clm}

\begin{figure}[!ht]
  \centering
    \includegraphics[width=.45\linewidth,trim={13cm 0 0 0},clip]{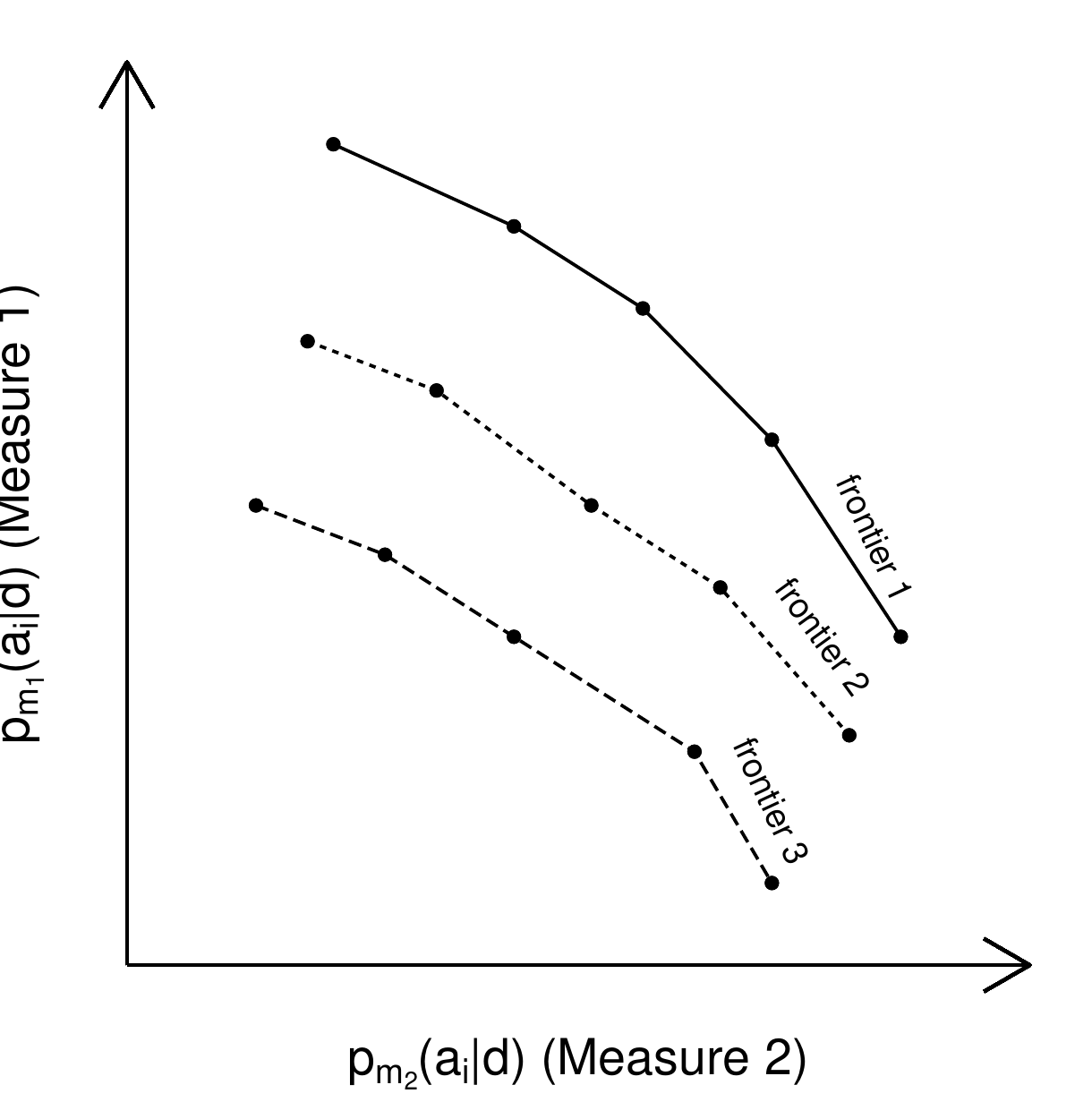}
      \caption{Dataset-Interest space.}
      \label{fig:ex_metatargets}
\end{figure}

To adapt the Pareto-Efficient Rankings to the multicriteria metatarget, we must first build the Dataset-Interest space $S_{d_i} = [p_{m_l} (a_j|d_i)]_{i=1}^{|A|}$. This space is shown in Figure~\ref{fig:ex_metatargets}, which shows two evaluation measures and fifteen algorithms as the axis and points of the problem, respectively. 

The Figure also shows the Pareto frontiers, which delimit the areas of Pareto dominance for the investigated algorithms, allowing to state when an algorithm is superior to another. The frontiers allow to understand two different relationships: algorithms within the same frontier can be considered similar, while those in different frontiers are effectively different. Similarly to the original work, the frontiers are calculated using the skyline operator algorithm~\citep{Lin2007}.

Formally, consider that the skyline operator creates a set $F$ of frontiers, where each frontier is represented as $f_k \in K$. This means that the output is now a sequence of pairs $\big(<a_j,f_k>\big)_{j=1}^{|A|}$, which assign each algorithm to a specific frontier. Our proposal at this point is to use such frontier values as algorithms rankings instead of using sorting mechanisms like previously. Considering how this procedure takes Pareto dominance into account the advantages are two-fold: (1) since we are not forced to assign a different ranking to all algorithms, this results in a more representative and fair assignment of algorithm ranking positions and (2) since the process is defined using a multidimensional Dataset-Interest space, then any number of evaluation measures $m_l$ can be used simultaneously.

\section{Empirical setup}\label{sec:setup}

This section presents the experimental setup. In order to ensure fair comparison of meta-approaches, based only on the predictive performance, the following constraints are adopted: (1) the baselevel datasets, algorithms and evaluation measures are exactly the same for all experiments; (2) all metalevel characteristics (multicriteria metatargets, algorithms and evaluation measures) are fixed, thus only the metafeatures change. Since the work considers baselevel and metalevel algorithms, we will refer to them as baselearners and metalearners, respectively.

\subsection{Collaborative Filtering}

The baselevel setup is concerned with the CF datasets, baselearners and measures used to evaluate the performance of CF baselearners when applied to these datasets. The 38 datasets used in the experiments are described in Table~\ref{tab:cf_data}, alongside a summary of their statistics, namely the number of users, items and ratings.

\begin{table}[!ht]	
\centering
\footnotesize
\caption{Summary description about the datasets used in the experimental study.} \label{tab:cf_data}
\begin{tabular}{l|c|c|c|l}
\hline
	Dataset & \#users & \#items & \#ratings & Reference\\ \hline
	Amazon Apps  & 132391 & 24366 & 264233 & \multirow{22}{*}{\citep{McAuley2013}}\\ \cline{1-4}
	Amazon Automotive  & 85142 & 73135 & 138039 \\ \cline{1-4}
	Amazon Baby  & 53188 & 23092 & 91468 \\ \cline{1-4}
	Amazon Beauty  & 121027 & 76253 & 202719 \\ \cline{1-4}
	Amazon CD  & 157862 & 151198 & 371275 \\ \cline{1-4}
	Amazon Clothes  & 311726 & 267503 & 574029 \\ \cline{1-4}
	Amazon Digital Music  & 47824 & 47313 & 83863 \\ \cline{1-4}
	Amazon Food  & 76844 & 51139 & 130235 \\ \cline{1-4}
	Amazon Games  & 82676 & 24600 & 133726 \\ \cline{1-4}
	Amazon Garden  & 71480 & 34004 & 99111 \\ \cline{1-4}
	Amazon Health  & 185112 & 84108 & 298802 \\ \cline{1-4}
	Amazon Home  & 251162 & 123878 & 425764 \\ \cline{1-4}
	Amazon Instant Video  & 42692 & 8882 & 58437 \\ \cline{1-4}
	Amazon Instruments  & 33922 & 22964 & 50394 \\ \cline{1-4}
	Amazon Kindle  & 137107 & 131122 & 308158 \\ \cline{1-4}
	Amazon Movies  & 7278 & 1847 & 11215 \\ \cline{1-4}
	Amazon Office  & 90932 & 39229 & 124095 \\ \cline{1-4}
	Amazon Pet Supplies  & 74099 & 33852 & 123236 \\ \cline{1-4}
	Amazon Phones  & 226105 & 91289 & 345285 \\ \cline{1-4}
	Amazon Sports  & 199052 & 127620 & 326941 \\ \cline{1-4}
	Amazon Tools  & 121248 & 73742 & 192015 \\ \cline{1-4}
	Amazon Toys  & 134291 & 94594 & 225670 \\ \cline{1-5}
	Bookcrossing  & 7780 & 29533 & 39944 & \multirow{1}{*}{\citep{Ziegler2005}}\\ \cline{1-5}
	Flixter  & 14761 & 22040 & 812930 & \multirow{1}{*}{\citep{Zafarani+Liu:2009}}\\ \cline{1-5}
	Jester1  & 2498 & 100 & 181560 & \multirow{3}{*}{\citep{Goldberg2001}}\\ \cline{1-4}
	Jester2  & 2350 & 100 & 169783 \\ \cline{1-4}
	Jester3  & 2493 & 96 & 61770 \\ \cline{1-5}
	Movielens 100k  & 94 & 1202 & 9759 & \multirow{5}{*}{\citep{GroupLens2016} }\\ \cline{1-4}
	Movielens 10m  & 6987 & 9814 & 1017159 \\ \cline{1-4}
	Movielens 1m  & 604 & 3421 & 106926 \\ \cline{1-4}
	Movielens 20m  & 13849 & 16680 & 2036552 \\ \cline{1-4}
	Movielens Latest  & 22906 & 17133 & 2111176 \\ \cline{1-5}
	MovieTweetings latest  & 3702 & 7358 & 39097 & \multirow{2}{*}{\citep{Dooms13crowdrec}}\\ \cline{1-4}
	MovieTweetings RecSys  & 2491 & 4754 & 20913 \\ \cline{1-5}
	Tripadvisor  & 77851 & 10590 & 151030 & \multirow{1}{*}{\citep{Wang2011} }\\ \cline{1-5}
	Yahoo! Movies  & 764 & 4078 & 22135 & \multirow{2}{*}{\citep{Yahoo} }\\ \cline{1-4}
	Yahoo! Music  & 613 & 4620 & 30852 \\ \cline{1-5}
	Yelp  & 55233 & 46045 & 211627 & \multirow{1}{*}{\citep{Yelp2016}}\\ \cline{1-5}
\end{tabular}
\end{table}

The experiments were carried out with MyMediaLite, a RS library~\citep{Gantner2011}. Two CF tasks were addressed: Rating Prediction and Item Recommendation. While the first aims to predict the rating an user would assign to a new instance, the second aims to recommend a ranked list of items. Since the tasks are different, so are the baselearners and evaluation measures required. 

The following CF baselearners were used for Rating Prediction: Matrix Factorization (MF), Biased MF (BMF)~\citep{Salakhutdinov2008}, Latent Feature Log Linear Model (LFLLM)~\citep{Menon2010}, SVD++~\citep{Koren2008}, 3 variants of Sigmoid Asymmetric Factor Model (SIAFM, SUAFM and SCAFM)~\citep{Paterek2007}, User Item Baseline (UIB)~\citep{Koren2010} and Global Average (GA). Regarding Item Recommendation, the baselearners used are BPRMF~\citep{Rendle2009}, Weighted BPRMF (WBPRMF)~\citep{Rendle2009}, Soft Margin Ranking MF (SMRMF)~\citep{Weimer2008}, WRMF~\citep{Hu2008a} and Most Popular (MP). All baselearners were selected since they represent different Matrix Factorization techniques, which are widely used in CF both in academia and industry due to their predictive power and computational efficiency.

In the experiments carried out, for Item Recommendation, the baselearners are evaluated using NDCG and AUC, while for Rating Prediction the evaluation measures NMAE and RMSE are used. All experiments are performed using 10-fold cross-validation. In order to prevent bias in favour of any baselearner, the hyperparameters were not optimised..

\subsection{Label Ranking as the Metalearning approach}

This work studies the performance of 4 meta-approaches: Rating Matrix metafeatures (RM)~\citep{Cunha2016}, Subsampling Landmarkers (SL)~\citep{Cunha2017}, the proposed Graph metafeatures (GR) and the Comprehensive metafeatures (CM). The last metafeatures are obtained aggregating all metafeatures from all existing meta-approaches and performing Correlation Feature Selection. Empirical validation has shown that setting the cutoff threshold at 70\% yields the best results. 

The multicriteria metatarget procedure is used to create the metatargets. Hence, for each CF problem studied (Rating Prediction and Item Recommendation), all specific evaluation measures are considered to create the Dataset-Interest spaces. This means that while NDCG and AUC are used for the Item Recommendation problem, NMAE and RMSE are used for Rating Prediction. Next, the Pareto-Efficient ranking procedure is employed for each dataset to generate a ranking of baselearners for Item Recommendation and another for Rating Prediction. The process is repeated for all remaining datasets in order to generate the complete metatargets. 

The MtL problem is addressed as a Label Ranking task. The following metalearners are used to induce metamodels: KNN~\citep{Soares2015}, Ranking Tree (RT), Ranking Random Forest(RF)~\citep{EXSY:EXSY12166}, and the baseline Average Ranking (AVG). The results are evaluated in terms of Kendall's Tau using leave one out cross-validation, due to the small number of meta-examples. Also, since we aim to obtain the best possible performance from the metalearners, we employ grid search optimisation. 

\section{Preliminary Analysis}\label{sec:preliminary}

\subsection{Graph Metafeatures Analysis}\label{sub:graph}

This analysis applies Correlation Feature Selection (CFS) to all proposed Graph metafeatures. It has two goals: (1) to remove unnecessary metafeatures and (2) to understand which levels of the proposed meta-approach are relevant to the investigated problem. Table~\ref{tab:graph_metafeatures} presents the metafeatures selected (65 out of 761), organised by level and number of metafeatures kept in the level. Each metafeature is presented using the notations introduced in Section~\ref{sec:main}.

\begin{table}[!ht]
\centering
\scriptsize
\caption{Graph metafeatures used in the experiments after CFS.}
\label{tab:graph_metafeatures}
\begin{tabular}{l|l|l}
\hline
Level    & Metafeatures selected & \# \\\hline
Graph    & none                  & 0 \\\hline
\multirow{3}{*}{Node}     & $\{G\}.\{authority,closeness\}.\{variance\}$ & 2 \\\cline{2-3}
         & $\{I\}.\{degree,diversity,eccentricity,PageRank\}.\{pf\}$ & 4 \\\cline{2-3}
         & $\{U\}.\{alpha,closeness,diversity\}.\{pf\}$ & 6 \\\hline
\multirow{4}{*}{Pairwise} & $\{I\}.\big[\{i_m/i_n\}.\{similarity\}.\{mpf\}\big].\{pf\}$  & 5  \\\cline{2-3}
         & $\{U\}.\big[\{u_k,u_l\}.\{similarity\}.\{variance\}\big].\{pf\}$  & 3  \\\cline{2-3}
         & $\{G\}.\big[\{g_i/g_j\}.\{similarity\}.\{variance\}\big].\{skewness\}$  & 1 \\\cline{2-3}
         & $\{I\}.\big[\{i_m/i_n\}.\{distances\}.\{sum\}\big].\{skewness\}$  & 1  \\\hline
\multirow{3}{*}{Subgraph} & $\makecell[l]{communities.\big[\{subgraph\}.\{alpha,authority,closeness,coreness,\\diversity,hub\ score, knn, strength\}.\{pf\}\big].\{pf\}}$ & 16 \\\cline{2-3}
         & $\makecell[l]{components.\big[\{subgraph\}.\{alpha,closeness,constraint,coreness,\\diversity,eccentricity,eigenvector\ centrality,knn\}.\{pf\}\big].\{pf\}}$ & 27	\\\hline
\end{tabular}
\end{table}

It can be seen that no Graph-level metafeatures are kept and only 12 out of 180 (7\%) Node-level are chosen. Regarding Pairwise and Subgraph metafeatures, 10 out 96 (10.4\%) and 43 out of 480 (9\%) metafeatures are selected, respectively. These results show the benefit of looking beyond standard Graph Theory measures, since a large amount of metafeatures kept by CFS belong to the metafeatures inspired by systematic and hierarchical decomposition procedures. 

\subsection{Comprehensive Metafeatures Analysis}\label{sub:comprehensive_metafeatures}

This section investigates the comprehensive metafeatures in order to understand how are they distributed across meta-approaches. The first step is to create a dataset with all preprocessed metafeatures from all meta-approaches. Next, CFS is applied to the dataset to select the most relevant metafeatures. Table~\ref{tab:comprehensive_meta} presents the metafeatures selected (the notations were introduced in Sections~\ref{sub:mtl_cf} and~\ref{sec:main}).

\begin{table}[!ht]
\centering
\scriptsize
\caption{Comprehensive metafeatures organised by meta-approaches.}
\label{tab:comprehensive_meta}
\begin{tabular}{l|c}
\hline
Meta-approach		 & Metafeatures \\\hline 
RM    	 & $\makecell[c]{
			I.\{count\}.\{gini,mode\} \\ I.\{mean\}.\{entropy,max,mode,sd\} \\ I.\{sum\}.\{kurtosis,mean\} \\  R.\{ratings\}.\{entropy,kurtosis,max,median,min,mode,skewness\} \\ U.\{count\}.\{kurtosis,max\} \\ U.\{mean\}.\{gini,kurtosis,mean,min,skewness\} \\ U.\{sum\}.\{entropy,mean,min,sd\} \\ nitems, nusers, nratings }$   \\\hline
SL     	 &  $\makecell[c]{ MostPopular.AUC , \\WBPRMF.\{AUC,NDCG\}, \\BMF.\{NMAE,RMSE\}, \\LFLLM.\{NMAE,RMSE\}, \\SCAFM.NMAE }$\\\hline
GR 		 &  $\makecell[c]{ G.\big[\{g_i/g_j\}.\{similarity\}.\{variance\}\big].\{skewness\} \\ communities.\big[\{subgraph\}.\{alpha,authority,coreness,\\diversity,hub\ score, knn\}.\{pf\}\big].\{pf\} \\ components.\big[\{subgraph\}.\{alpha,closeness,constraint,coreness,\\diversity,eigenvector\ centrality\}.\{pf\}\big].\{pf\} \\ \{I\}.\big[\{i_m/i_n\}.\{distances,similarity\}.\{count,sum\}\big].\{skewness\} \\ \{U\}.\big[\{u_k,u_l\}.\{similarity\}.\{variance\}\big].\{skewness\} \\ \{I\}.\{eccentricity\}.\{skewness\} \\ \{U\}.\{alpha\}.\{skewness\}}$\\\hline       
\end{tabular}
\end{table}

It can be observed that the comprehensive meta-approach contains metafeatures from all meta-approaches, showing that they are complementary. However, different numbers of metafeatures are selected from each meta-approach: RM and GR have the highest contribution, with 29 and 26 metafeatures, respectively. The least contributing meta-approach, SL, provides only 8 metafeatures. Notice that SL metafeatures depend on the metatarget (since they are bounded by the baselearners and evaluation measures), while RM and GR are metatarget independent. 

\subsection{Multicriteria Metatarget Analysis}\label{sub:multicriteria_metatarget}

\begin{figure}[!ht]
  \centering
    \includegraphics[width=.8\linewidth]{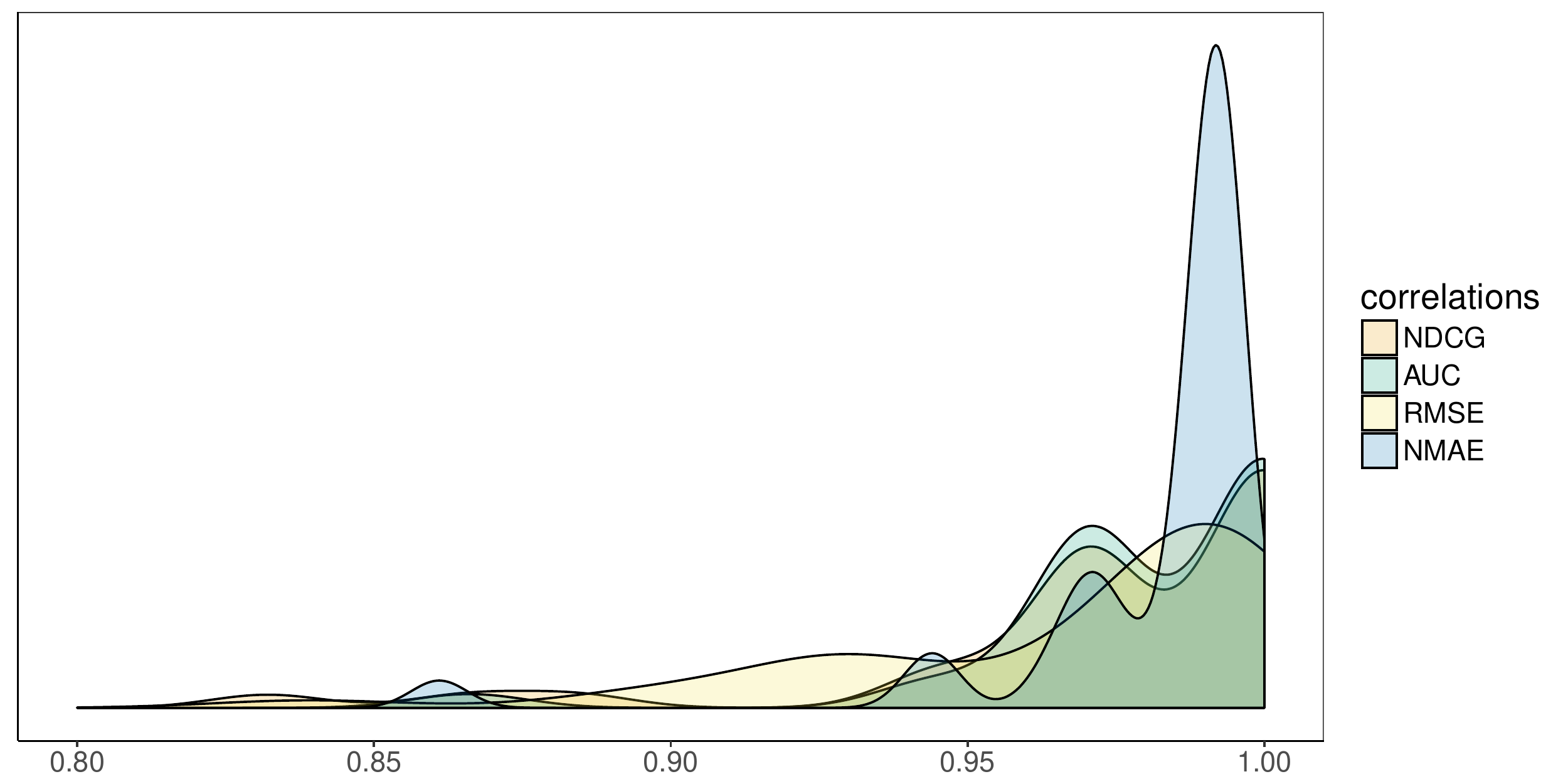}
      \caption{Distributions of correlations between individual and multicriteria rankings.}
      \label{fig:ranking_multicriteria_dist}
\end{figure}

The last analysis validates the multicriteria metatarget methodology. For such, it is important to understand how aligned are the multicriteria metatargets with the individual metatargets. This is crucial since we want to assure the main trends in the individual metatargets rankings are not completely lost. Thus, while some deviations in ranking positions are accepted (and in fact expected due to the heterogeneity of evaluation measures used), completely different rankings should not be produced. If they are, they do not reflect the individual realities observed, making it difficult to assess that the procedure works properly. To verify if this occurs, the following procedure is used: (1) calculate the metatargets for each baselevel dataset independently, using the multicriteria metatarget strategy and (2) calculate the correlation between the rankings of baselearners produced by each individual metatarget and those in the multicriteria metatarget. Figure~\ref{fig:ranking_multicriteria_dist} illustrates the distributions of correlations. The results are zoomed in the $[0.8,1]$ range, which contains over 93\% of the correlations.  

According to these results, most correlations fall in the $[0.9,1]$ range, indicating that the individual and multicriteria metatargets are very similar. Therefore, the multicriteria metatarget strategy proposed was successful. However, we must also understand what happened in the least successful cases. Table~\ref{tab:correlations} details these correlations. In it, the dataset, the measures used to create the original rankings and the respective correlations to the multicriteria metatargets are presented.

\begin{table}[!ht]
\centering
\scriptsize
\caption{Correlations between individual and multicriteria metatargets with threshold below 80\%. M1 and M2 refer to the first and second evaluation measures.}
\label{tab:correlations}
\begin{tabular}{l|l|l|l|l}
\hline
	Dataset & M1 & M2 & corr(M1, Multicriteria) & corr(M2, Multicriteria) \\ \hline
	Amazon Movies & NDCG & AUC & 0.884 & -0.177 \\ \hline
	Flixter & NDCG & AUC &  0.866 &  0.289 \\ \hline
	Movielens Latest & NDCG & AUC &  0.289 &  0.866 \\ \hline
	MovieTweetings Latest & NDCG & AUC &  0.949 &  0.474 \\ \hline
	Tripadvisor & NDCG & AUC &  0.667 &  0.000 \\ \hline
	Jester1 & RMSE & NMAE & 0.935 & 0.355 \\ \hline
	Jester2 & RMSE & NMAE &  0.935 & 0.355 \\ \hline
	Jester3 & RMSE & NMAE & 0.936 & 0.484 \\ \hline
	Movielens 10m & RMSE & NMAE &  0.738 & 0.975 \\ \hline
\end{tabular}
\end{table}

According to this table, the cases where the correlation between the multicriteria metatarget and an original ranking falls below 80\% only occur for a single measure. Thus, in these cases, the differences in performance in one measure overpower the performance differences in the other. The only exception is the Tripadvisor dataset, used in the Item Recommendation task. In this case, every baselearner receives a ranking value of 1 or 2, since there is a high discrepancy in the rankings created by each evaluation measure. 

\section{Experimental Results}\label{sec:results}

\subsection{Metalevel accuracy}

\begin{figure}[!ht]
    \centering
    \begin{subfigure}[b]{0.7\textwidth}
        \includegraphics[width=\textwidth]{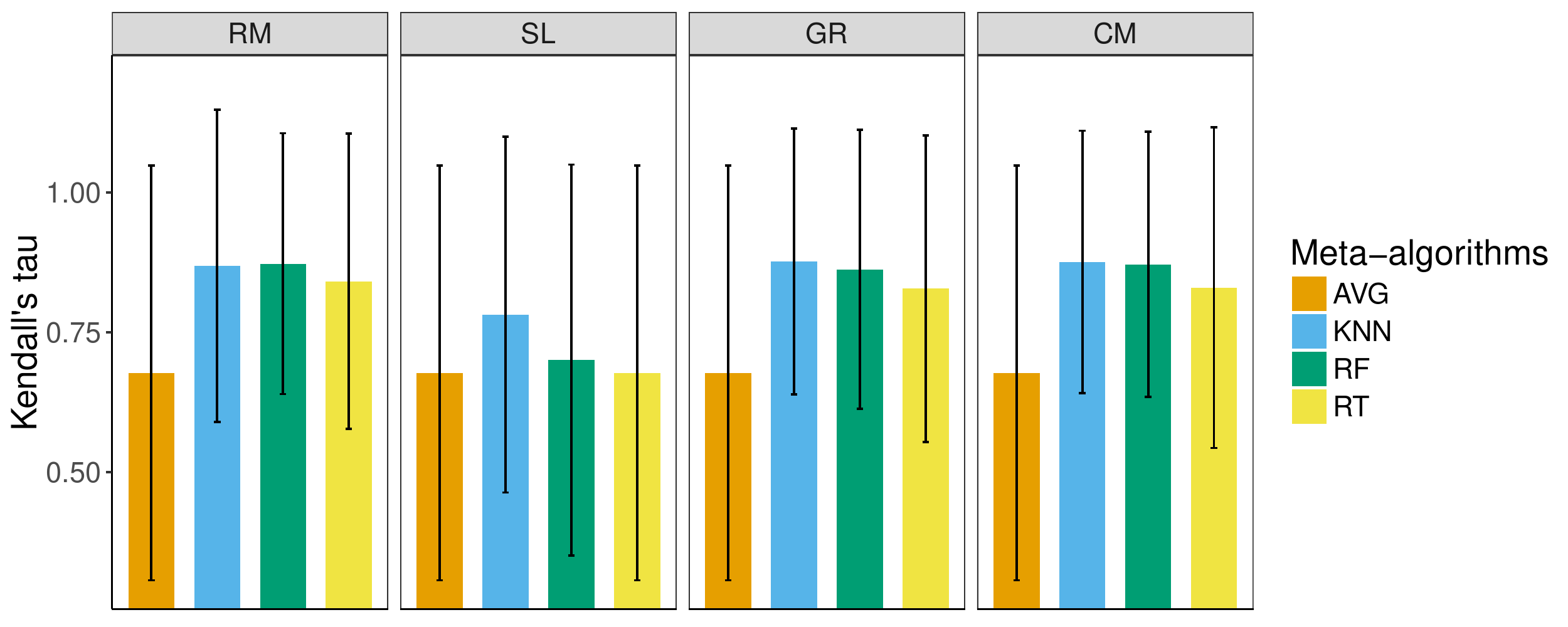}
        \caption{Item Recommendation scope.}
        \label{fig:meta_eval_IR}
    \end{subfigure}
    ~ 
    \begin{subfigure}[b]{0.7\textwidth}
        \includegraphics[width=\textwidth]{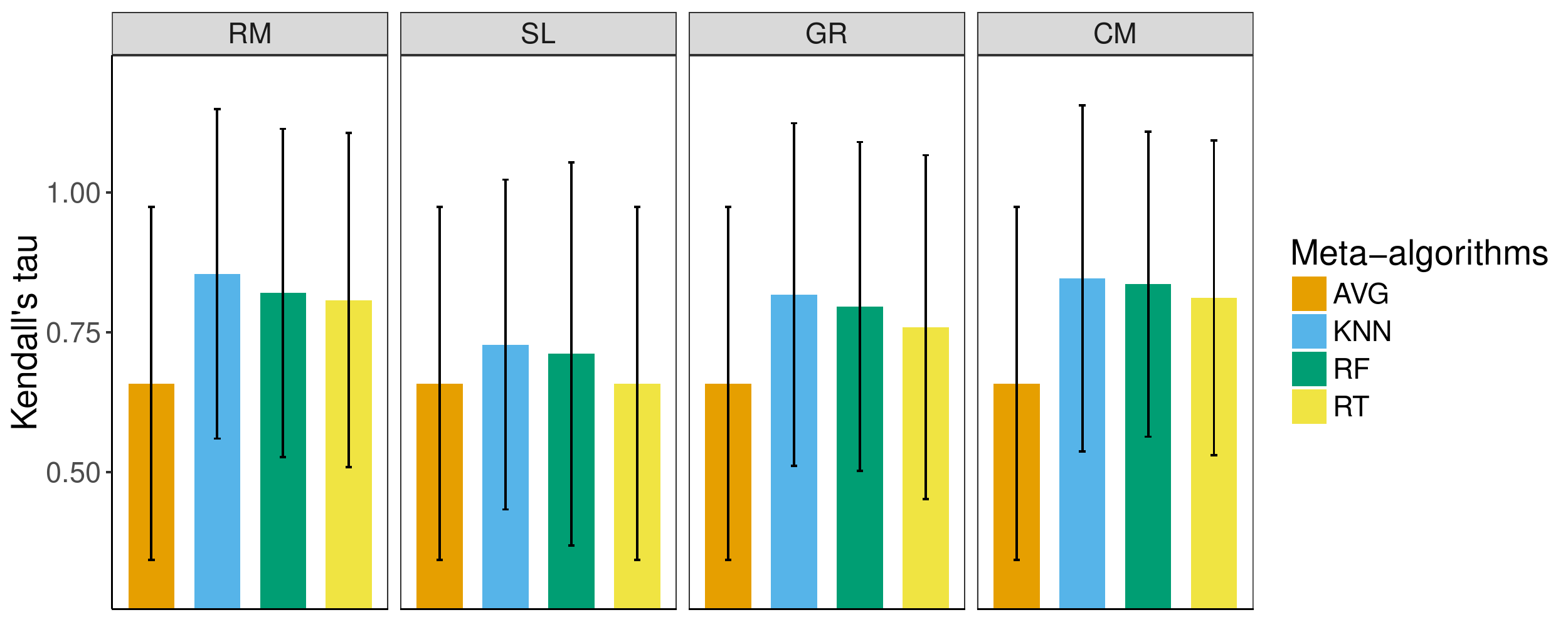}
        \caption{Rating Prediction scope.}
        \label{fig:meta_eval_RP}
    \end{subfigure}
    \caption{Overall ranking accuracy aggregated by meta-approaches and metalearners.}\label{fig:meta_eval}
\end{figure}

Metalevel performance can be measured by the accuracy of the metalevel prediction, which evaluates whether the best ranking of baselearners is selected. To do so, the validation strategy compares, for each dataset, the predicted ranking with the true ranking. The ranking accuracy is measured using Kendall's tau coefficient. Figures~\ref{fig:meta_eval_IR} and~\ref{fig:meta_eval_RP} presents the average and standard deviation of the Kendall's tau coefficient for each metalearner and meta-approaches in both CF tasks studied.

According to the results, several observations can be made:
\begin{itemize}
\item \textbf{All metalearners are able to beat the baseline.} This is a specially important result since it means that all meta-approaches proposed are suitable for the CF algorithm selection problem. 
\item \textbf{GR is competitive to RM.} the results show that in Item Recommendation it is slightly better, while in Rating Prediction it is slightly worse. Although we hoped that the graph metafeatures would be better than the related work meta-approaches, it is still relevant that they are a suitable alternative. 
\item \textbf{SL is always the worst meta-approach.} Since this has happened in previous work~\citep{Cunha2017}, it is not a surprise. 
\item \textbf{CM metafeatures do not perform better than the remaining meta-approaches.} These results show that no gain was obtained by using metafeatures from different domains. However, since the ranking accuracy scores were already high, close to 90\%, the difficulty to increase the predictive performance is understandable. 
\item \textbf{Metalearners are ranked in the following order:} KNN is usually the best, followed by RF and RT in second and third places, respectively. 
\end{itemize}

\begin{figure}[!ht]
    \centering
    \begin{subfigure}[b]{0.4\textwidth}
        \includegraphics[width=\textwidth,trim={5.5cm 2.5cm 4.5cm 2cm},clip]{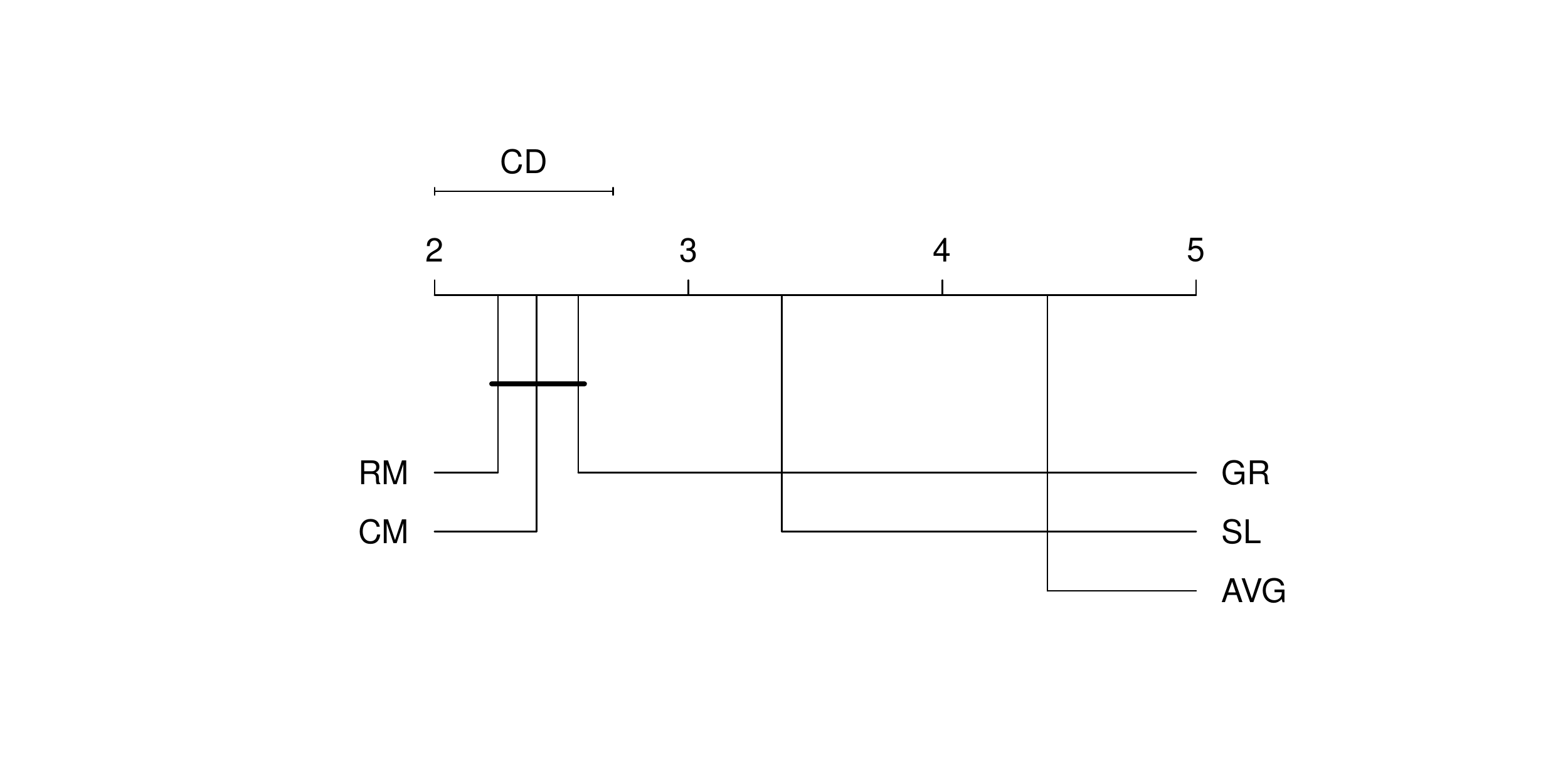}
        \caption{Meta-approaches.}
        \label{fig:cd_meta_approach}
    \end{subfigure}
    ~ 
    ~
    ~ 
    \begin{subfigure}[b]{0.4\textwidth}
        \includegraphics[width=\textwidth,trim={5.5cm 2.5cm 4.5cm 2cm},clip]{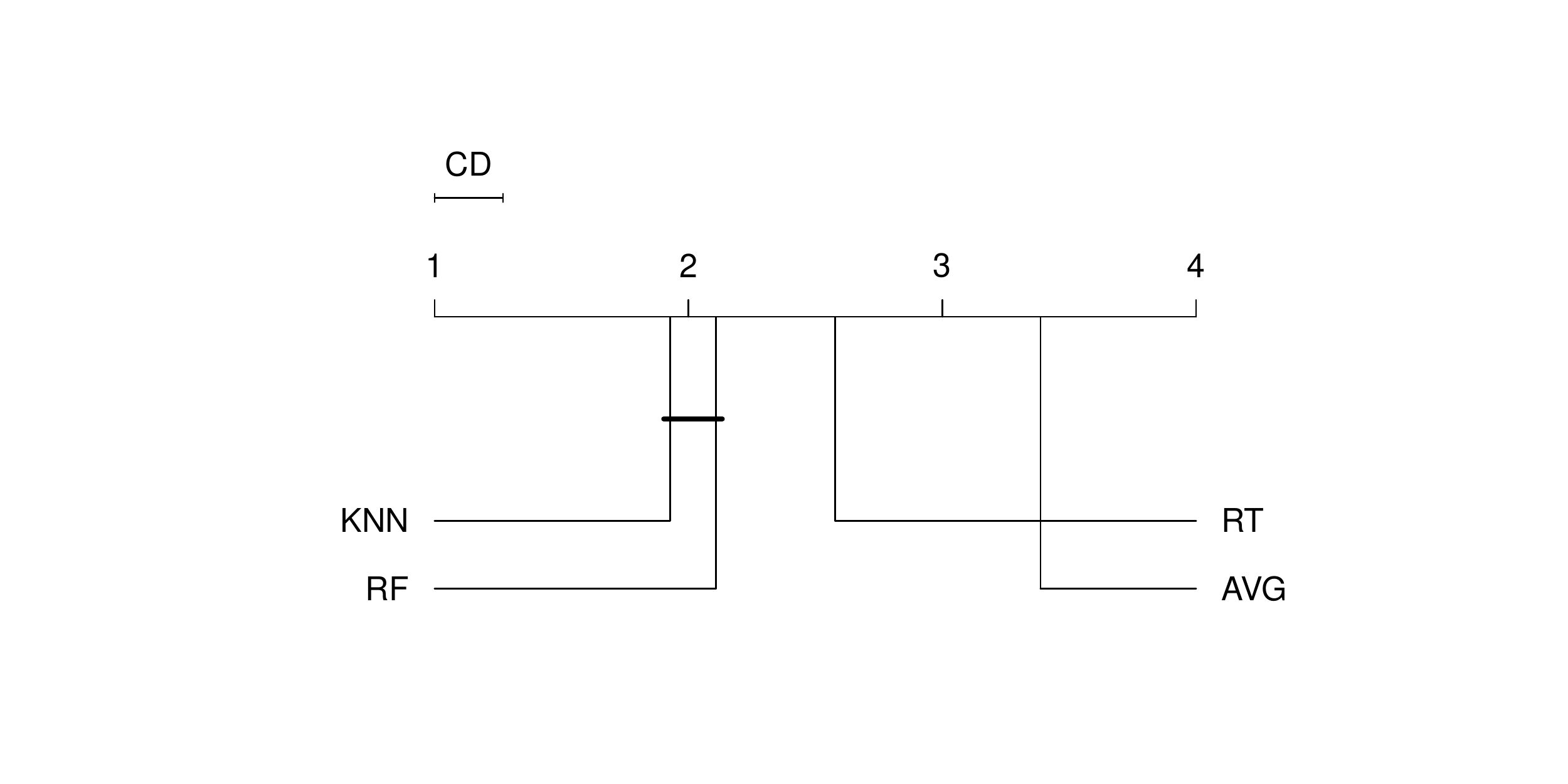}
        \caption{Metalearners.}
        \label{fig:cd_algs}
    \end{subfigure}
    \caption{Critical Difference diagrams.}\label{fig:cd_diagrams}
\end{figure}

To evaluate the statistical significance of the differences in the performance results, Critical Difference (CD) diagrams~\citep{Demsar2006} were used. These diagrams are used here to rank several metalearners based on their performance on several datasets. Furthermore, they include a measure of statistical significance - CD interval - which is represented by the CD line. Thus, when two metalearners are connected by this line, there is no statistically significant difference between them. These diagrams were used to to assess two different perspectives: 

\begin{itemize}
\item \textbf{Meta-approach:} to see whether there are differences among meta-approaches, the best performing metalearner per meta-approach was selected (KNN was chosen for all meta-approaches except GR, which uses RFR). This means that its Kendall's Tau performances for all datasets are used as the meta-approach representation. The results are presented in Figure~\ref{fig:cd_meta_approach}.
\item \textbf{Metalearners:} to assess how different are the metalearners on a global perspective, the CD methodology was applied to all available performances. The results are presented in Figure~\ref{fig:cd_algs}.
\end{itemize} 

These results validate all previous observations and show that there is no statistically significant difference between RM, CM and GR. This proves that both meta-approaches, although ranked slightly worse, are as good as the best CF metafeatures. In terms of metalearners, the results show that there is no statistically significant difference between KNN and RF, although both are better than RT. Considering that RF requires more computational resources to train and predict, KNN seems to offer the best solution.

\subsection{Impact on the baselevel performance}

Beyond understanding whether the rankings of algorithms predicted are correct, one must understand how costly are when the rankings are incorrect. This is important because a switch of 2 algorithms in a ranking may be less penalized in the metalevel accuracy, but it may be catastrophic in terms of baselevel performance. 

Hence, this strategy needs to compare metalearners by baselevel performances of their predicted rankings of baselearners. We consider each ranking position independently (defined by a threshold $t$) in order to analyse each algorithm independently. Furthermore, since we are evaluating the thresholds from the best to the worst algorithm (i.e. following the ranking order), we must assure that at any threshold $t$ we have the best possible performance from previous thresholds. The consequence is that at each threshold $t$, the performance will be either better than or equal to the previous threshold. This is necessary to ensure that the metamodel is judged by the best performance possible. The metamodel is then represented as a sequence of baselevel performance values with constant or increasing values. The procedure to obtain the data for this analysis is the following: 

\begin{itemize}
\item For a dataset $d_i$, consider the best ranking of algorithms $R_{d_i}$. This ranking is directly represented by a performance vector $P_{d_i}$.
\item Consider now a predicted ranking $\hat{R_{d_i}}$ provided by a metamodel for $d_i$. 
\item The predicted performance vector $\hat{P}_{d_i}$ is created by obtaining the baselevel performances of every algorithm $\hat{a_i}$ from the original performances $P_{d_i}$. To do so, the algorithms from $\hat{R_{d_i}}$ and $R_{d_i}$ are matched by name.
\item The performance vector $\hat{P}_{d_i}$ is then regularized to ensure that at each threshold $t$ the values are set to be either better or the same as the previous threshold.
\item The process is repeated for all datasets, obtaining then a set of performance vectors. Then, the performance values are averaged for each threshold $t$, creating an average performance of the metamodel in terms of baselevel performance.
\end{itemize}

\begin{figure}[!ht]
    \centering
    \begin{subfigure}[b]{0.9\textwidth}
        \includegraphics[width=\textwidth]{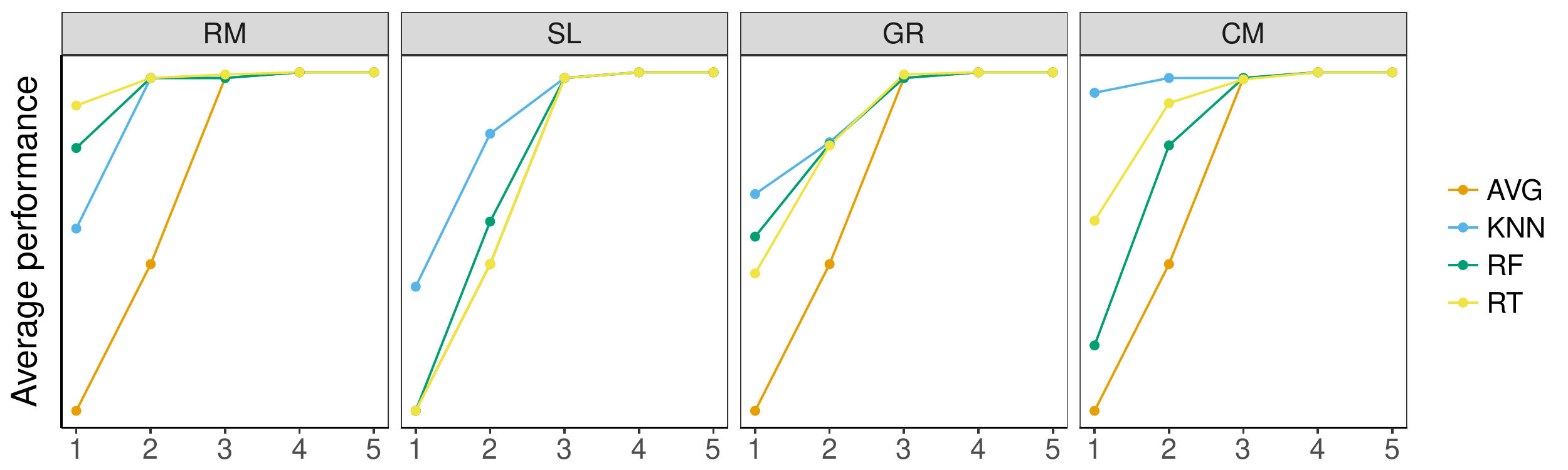}
        \caption{Item Recommendation scope.}
        \label{fig:base_IR}
    \end{subfigure}
    ~ 
    \begin{subfigure}[b]{0.9\textwidth}
        \includegraphics[width=\textwidth]{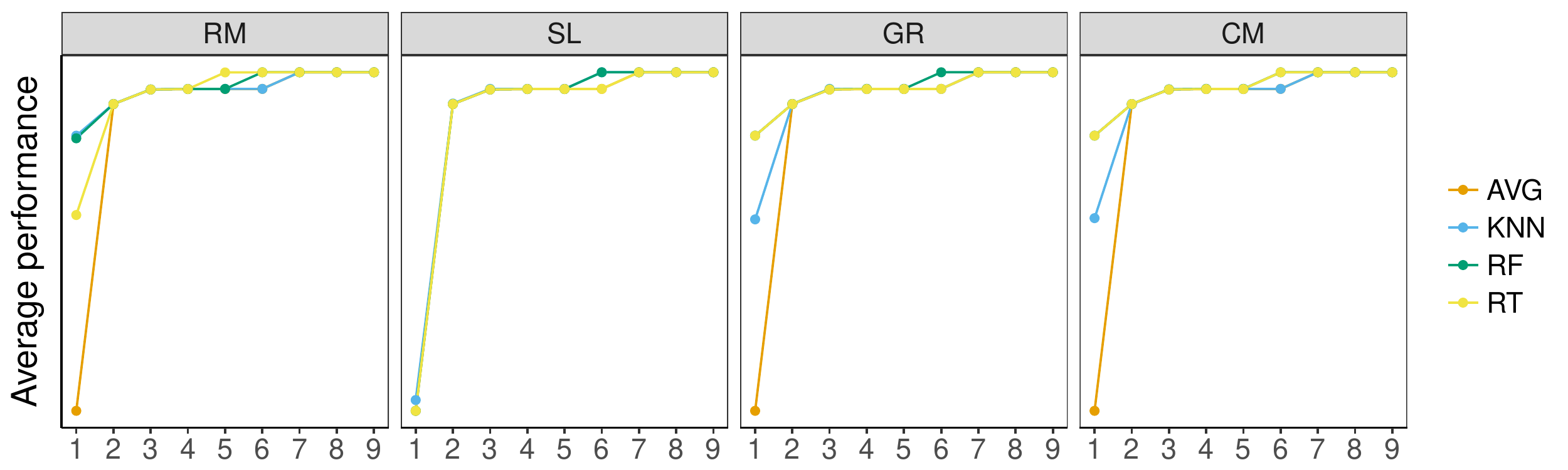}
        \caption{Rating Prediction scope.}
        \label{fig:base_RP}
    \end{subfigure}
    \caption{Impact on the baselevel performance.}\label{fig:base}
\end{figure}

Figures~\ref{fig:base_IR} and~\ref{fig:base_RP} present these results, aggregated by meta-approach. 

According to the results, a metalearner is usually better than the baseline for all meta-approaches in both scopes (except for SL case in Rating Prediction). This happens for $t \in \{1,2\}$ and $t=1$ in Item Recommendation and Rating Prediction, respectively. Therefore, the predictions resulted in better baselevel performances for higher positions in the predicted rankings of baselearners than the baseline. Given that practitioners pay more attention to the highest ranked predicted baselearners, these results are very important.

Regarding the metalearners, only KNN constantly outperforms the baseline and in Item Recommendation it is the best choice for all meta-approaches, except RM. In Rating Prediction, however, RF and RT tend to work better, except for the RM case. Considering also the meta-accuracy results, although KNN is more accurate in predicting rankings, some of its mistakes have a high cost. This is clear in Rating Prediction, although it also happens in Item Recommendation. 

\subsection{Metaknowledge analysis}

This metaknowledge analysis looks for the most important metafeatures by meta-approach. Since there is no standardised feature importance procedure for Label Ranking, a heuristic strategy is used~\citep{Cunha2018}: to traverse all trees in Random Forest metamodels, assign all metafeatures with the respective tree level (i.e. its ranking in the specific tree) and average the results per metafeature. The final scores indicate the metafeature's global ranking. Table~\ref{tab:feature_imp} shows the top-10 metafeatures for all meta-approaches using this method. 

\begin{table}[!ht]	
\centering
\scriptsize
\caption{Ranking of features per meta-approach. The following names are abbreviated: 'entropy' (ent), 'kurtosis' (kur), 'skewness' (ske), 'variance' (var), 'similarity' (sim), 'coreness' (core), 'communities' ('com') and 'components' ('comp').} \label{tab:feature_imp}
\begin{tabular}{r|c|c|c|c}
\hline
	 & RM & SL & GR & CM \\ \hline
	1 & I.count.ent & WBPRMF.AUC & G.[\{$g_i/g_j$\}.sim.var].ske & G.[\{$g_i/g_j$\}.sim.var].ske \\ \hline
	2 & I.count.kur & WBPRMF.NDCG & com.[sub.diversity.var].ske & com.[sub.core.ent].ske \\ \hline
	3 & I.count.mean & BPRMF.NDCG & com.[sub.alpha.var].var & LFLLM.NMAE \\ \hline
	4 & I.count.max & BMF.NMAE & I.[\{$i_m/i_n$\}.sim.sum].var & BMF.RMSE \\ \hline
	5 & U.count.mean & MP.AUC & com.[sub.alpha.mean].ent & BMF.NMAE \\ \hline
	6 & nusers & LFLLM.NMAE & comp.[sub.closeness.ske].ske & R.ratings.ske \\ \hline
	7 & I.mean.ske & LFLLM.RMSE & com.[sub.core.ent].ske & U.sum.mean \\ \hline
	8 & I.count.gini & UIB.RMSE & I.[\{$i_m/i_n$\}.sim.var].ske & com.[sub.alpha.mean].ske \\ \hline
	9 & I.mean.ent & SIAFM.NMAE & comp.[sub.alpha.ske].ske & LFLLM.RMSE \\ \hline
	10 & R.rat.ske & SCAFM.NMAE & com.[sub.alpha.mean].ske & R.ratings.mode \\ \hline
\end{tabular}
\end{table}

The results allow to make several observations, which are presented next, organised by meta-approach.

\subsubsection{Rating Matrix metafeatures (RM)}

Analysing the most important RM metafeatures, it is possible to observe that:

\begin{itemize}
\item The most important object is the $item$ (used in 7 metafeatures). The remaining metafeatures are equally distributed by $rating\ matrix$, $users$ and a generic metafeature (i.e. $nusers$);
\item The most meaningful function is the $count$ (present in 6 metafeatures). Besides, functions $mean$ and $ratings$ appear twice and once, respectively and no metafeature in the top-10 uses the $sum$ function;
\item In terms of post-functions, the results are evenly distributed: $entropy$ (2), $mean$ (2), $kurtosis$ (1), $skewness$ (2), $maximum$ (1) and $gini$ (1). Thus, $minimum$, $standard\ deviation$ and $mode$ are left out of this ranking.
\end{itemize}

In summary, the most important RM metafeatures are related to statistics from the distribution of the number of ratings per item (i.e. $I.count.*$). This happens because these metafeatures appear in the top-4 metafeatures but also because the object $i$ and function $count$ are the most frequent in this ranking. The authors believe that these are the most informative metafeatures, since they allow to discern between baselearners with or without bias towards popular items. For instance, these metafeatures allow to decide when the baseline $MostPopular$ is the best baselearner or whether more intelligent baselearners should be used.

\subsubsection{Subsampling Landmarkers metafeatures (SL)}

Regarding SL, several observations regarding which baselearners and evaluation measures are represented in the top metafeatures can be made:

\begin{itemize}
\item In the Item Recommendation task, the following baselearners are considered in the best ranked metafeatures: BPRMF, WBPRMF and MP. In fact, the top-2 metafeatures in this case refer to the WBPRMF baselearner. However, no metafeature using the SMRMF and WRMF baselearners is available;
\item Regarding the Rating Prediction task, the baselearners available are almost evenly distributed: LFLLM (2), BMF (1), UIB (1), SIAFM (1) and SCAFM (1). However, 4 baselearners (MF, SVD++, SUAFM and GA) are not present. 
\item All evaluation measures are present in the ranking, although with different frequencies: NDCG (2), AUC (2), NMAE (4) and RMSE (2). 
The fact that these results are biased towards the Rating Prediction task is expected, given the fact that there are more landmarkers in Rating Prediction than in Item Recommendation (i.e. 18 versus 10).
\end{itemize}

These results show that the SL metafeatures are well distributed between CF tasks, both in terms of baselearners and evaluation measures. They also shown an interesting pattern: 4 out of the top-5 metafeatures are related to the Item Recommendation task and all remaining metafeatures are related to the Rating Prediction task. Therefore, the results seem to point out that these metafeatures are most informative for Item Recommendation than to Rating Prediction. 

\subsubsection{Graph metafeatures (GR)}

For the newly proposed GR metafeatures, the analysis shows that:

\begin{itemize}
\item Most objects available in the top-10 metafeatures are related to the $communities$ and $components$, with 7 out of 10 metafeatures. Only two refer to the item graph $I$ and one to the original graph $G$. Also, no metafeature here represents the user graph $U$. 
\item Considering the wide range of Graph Theory functions used, the most important are $alpha$ and $similarity$ with 4 and 3 metafeatures, respectively. The remaining functions are evenly distributed: $diversity$ (1), $closeness$ (1) and $coreness$ (1). The functions $authority$, $degree$, $eccentricity$, $hub score$, $eigenvector\ centrality$, $knn$, $neighbours$, $local\ scan$, $PageRank$ and $strength$ do not appear in any of the top-10 metafeatures.
\item Regarding OL post-functions, there is a clear trend towards $skewness$, with 7 related metafeatures. The remaining post-functions are $variance$ (2) and $entropy$ (1). Notice that in this level, the function $mean$ is not in the top-10.
\item The IL post-functions (either $pf$ or $mpf$) are evenly distributed: $variance$ (4), $skewness$ (2), $mean$ (2), $sum$ (1) and $entropy$ (1). 
\end{itemize}

Overall, the results show: (1) pairwise- and subgraph-levels are the most informative metafeatures, since they are the only metafeatures to appear in the top-10 metafeatures and (2) the most informative metafeatures belong to the subgraph-level, having 7 in the top-10. Hence, this analysis confirmed the benefits of exploring metafeatures beyond standard Graph Theory measures, namely by taking advantage of hierarchical and systematic procedures to generate metafeatures.  

\subsubsection{Comprehensive metafeatures (CM)}

Regarding the CM metafeatures, it was seen that:

\begin{itemize}
\item The metafeatures are evenly distributed among the meta-approaches: RM (3), SL (4), and GR (3). These results show that CM gives more importance to the metafeatures created with the meta-approach with worst performance: SL;
\item The top-5 metafeatures belong to GR and SL meta-approaches, with the first and second belonging to GR and the remaining assigned to SL. Metafeatures from the RM meta-approach are available only in the sixth and tenth position;
\item The most important metafeatures per meta-approach are not necessarily the best for CM: although the 3 GR metafeatures in CM are also available in GR's scope in the same ranking order, the same does not happen for the others. Specifically, only the worst RM metafeature is chosen, while SL is only represented by BMF and LFLLM. 
\item The opposite to the previous situation is also true: the best ranking metafeatures per meta-approach in the CM do not appear in the individual case. 
\end{itemize}

In summary, the results show that the CM metafeatures do not always consider the best performing metafeatures by meta-approach to create the overall representation. Several reasons may justify this behaviour, which include (1) either combination of several domains drastically changes the dimensions of the feature space and, with it, the data patterns which Label Rankings metalearners take advantage of or (2) the methodology used to select the metafeatures should be replaced. If this proves to be true, it may be the reason why the CM metafeatures performed worse than the other meta-approaches.

Lastly, it is important to notice that although the authors have tried to infer why the observed patterns have occurred, the reasons for most observations are still unknown. This happens because algorithm selection is not a trivial task, especially when we consider the complexity of metafeatures and metatargets used here. Hence, future work is necessary to give meaning to the patterns found here. 

\section{Conclusions}\label{sec:conclusions} 

This work provides the most comprehensive study of Collaborative Filtering algorithm selection to date. It does so by introducing two new meta-approaches (graph and comprehensive) and a procedure to create multicriteria metatargets. Also, it uses a large and controlled experimental procedure to properly compare a large variety of metafeatures. Furthermore, it performs an extensive metaknowledge analysis, which yields the best performing metafeatures. Now we provide the main conclusions of this work and directions for future work for several topics.

\paragraph{\textbf{Graph metafeatures: } This meta-approach has shown comparable results to those of state of the art metafeatures, even despite showing slightly worse performance in terms of impact on the baselevel performance. Inspecting the metaknowledge allows to understand that pairwise and subgraphs levels are the most meaningful to the problem. These results are particularly important since they validate our motivation to model Collaborative Filtering as a graph and to take advantage of a hierarchical and systematic approach to design metafeatures. Future work tasks must include the extension of the experimental procedure.}

\paragraph{\textbf{Comprehensive metafeatures: } These metafeatures have shown to be complementary since they include metafeatures from all available meta-approaches. Despite the high expectations set to this meta-approach, the experimental results have shown that its performance is no statistically significantly better than the state of the art metafeatures. Hence, it seems that little can be gained from using the comprehensive metafeatures. Inspecting the metaknowledge allows to perceive that not always the best individual metafeatures are selected. Hence, this points out to the idea that more advanced feature selection techniques should be used.}

\paragraph{\textbf{Multicriteria metatargets: } The evaluation of the metatarget methodology showed its effectiveness to create rankings of algorithms from multiple evaluation measures. This has been proven since in the vast majority of cases the correlation between individual and multicriteria metatargets is above 80\%. Also, since the experimental performance of Rating Matrix and Subsampling Landmarkers metafeatures are consistent with those presented in the related work, it also validates the multicriteria metatargets procedure proposed. However, there are still at least two important future work tasks to be addressed: to validate this methodology using more than two evaluation measures per scope and to apply the multicriteria metatarget technique to Metalearning tasks in other domains. }

\paragraph{\textbf{Metaknowledge: } This work has also presented important conclusions regarding the importance of metafeatures: while Rating Matrix metafeatures give more importance to statistics from the distribution of the number of ratings per item, Subsampling Landmarkers seem to be more important to the Item Recommendation task rather than the Rating Prediction one. However, the main issue found in the metaknowledge analysis lies in the fact that comprehensive metafeatures seem to ignore the best metafeatures from individual meta-approaches. This is makes it particularly important to understand why the CF metafeatures presented are important to the Collaborative Filtering problem.}

\bibliographystyle{spbasic}      
\bibliography{myrefs}   

\end{document}